\documentclass[12pt]{iopart}
\usepackage[latin1]{inputenc}
\usepackage{amsfonts}
\usepackage{amssymb}
\usepackage{graphicx}
\begin{document}
\title{Ising model with memory: coarsening and persistence properties}
\author{F Caccioli$^{1,2}$, S Franz$^3$ and M Marsili$^4$}
\address{$^1$ International School for Advanced Studies, via Beirut 2-4, 34014 Trieste (Italy)}
\address{$^2$Istituto Nazionale di Fisica Nucleare, sezione di Trieste}
\address{$^3$Universit\'{e} Paris-Sud, LPTMS, UMR8626,
B\^{a}t. 100, 91405 Orsay cedex (France)}
\address{$^4$ The Abdus Salam International Centre for Theoretical Physics, Strada Costiera 14,
34014 Trieste (Italy)}
\eads{\mailto{caccioli@sissa.it},  \mailto{silvio.franz@lptms.u-psud.fr}, \mailto{marsili@ictp.it}}

\begin{abstract}We consider the coarsening properties of a kinetic Ising model with a memory field. The probability of a spin-flip depends on the persistence time of the spin in a state. The more a spin has been in a given state, the less the spin-flip probability  is. We numerically studied the growth and persistence properties of such a system on a two dimensional square lattice. The memory introduces energy barriers which freeze the system at zero temperature. At finite temperature we can observe an apparent arrest of coarsening for low temperature and long memory length. However, since the energy barriers introduced by memory are due to local effects, there exists a timescale on which coarsening takes place as for the Ising model. Moreover the two point correlation functions of the Ising model with and without memory are the same, indicating that they belong to the same universality class.
\end{abstract}
\section{Introduction}
Coarsening and persistence properties in ferromagnetic systems following a quench from a disordered into an ordered phase have been widely studied in the past decades (\cite{Bra94}, \cite{Hum91}, \cite{Der97}, \cite{Hin97}, \cite {Der94}, \cite{Sta94}, \cite{Der96}).  These systems do not order instantaneously, but the length scale of ordered regions grows in time as different phases compete to select the equilibrium state. In the thermodynamic limit final equilibrium is not achieved, but a scaling regime characterized by a unique typical length is reached at long times. In this regime the statistics of domains is the same at all times apart from a global change of scale.  Different  growth kinetics can arise depending on quite general properties of the systems. For the two dimensional (kinetic) Ising model it is well known that the characteristic length grows asymptotically as
\begin{equation}
L(t)\sim t^{1/2},
\end{equation}
as predicted by the Allen-Cahn equation for systems with non conserved order parameter \cite{Bra94}.
The fraction of persistent spins ($r(t)$) (to be more precise the fraction of space which remains in the same phase up to time $t$ \cite{Der97}) decays instead as
 \begin{equation}
r(t)\sim t^{-\theta},
\end{equation}
with $\theta\sim0.22$ independent of the temperature (\cite{Der97}, \cite{Hin97}).
In this paper we consider a kinetic Ising model in which we introduce a field which breaks detailed balance. The field makes sure that the more a spin remains in a state the less is the probability for that spin to flip and it has a natural interpretation in terms of "memory of spins".
Apart from its intrinsic interest, the motivation for introducing  such a term comes from the possible applications in the context of
opinion dynamics, where Ising like models have been often considered
 (\cite{Gal97}, \cite{Cas07}, \cite{Mic05}, \cite{Bor06}). In a statistical physics approach an opinion
is usually model by a spin variable, which represents the position of an agent with respect to a specific issue
(e.g. yes/no, Linux/ Windows, democratic/republican). The aim is to understand the collective behavior of a
society, depending on the "interaction" among spins which models the way in which opinions of agents form and
influence those of other agents. Random fields on each site have also been introduced to represents the personal
orientation of each agent in absence of social interactions. In such a social interpretation of the Random Field
Ising Model (RFIM) \cite{Gal97}, each agent has to make a compromise between her own a priori belief (the random
field) and the stimuli due to social interactions (nearest-neighbor interactions). This gives rise to very
interesting domain growth properties. When people have to make a choice, however, usually they refer to their
own past experience. The fact that people can learn from their past is not taken into account in an RFIM. It is
then interesting to introduce in these models a field which represents the agents' memory. This field should
depend on the history so to mimic the learning process of agents. An interesting question is how this learning
process can influence the kinetics of domains. We consider here a very simple
memory field, which is self-generated and develops spatial correlations.  We study the growth and persistence properties of such a system
on a two dimensional square lattice. We show that, even though there is a pinning action of spins that leads to
a freezing at zero temperature, at finite temperature the system is in the same universality class of the Ising
model. This is due to the fact that the energy barriers introduced by memory are local and do not depend on the
domain size, as it happens instead for the RFIM.

\section{The model}
We consider a system with $N$ spins $\sigma_i=\pm 1$ . At time $t$ we assign to each spin an effective
energy given by
\begin{equation}
\label{energy} E_i(t)=-\sigma_i(t)\sum_j\sigma_j(t)-j_{mem}(1-\delta)\sigma_i(t)\phi_i(t-1),
\end{equation}
where the sum running over $j$ is  a sum over the nearest neighbors of  spin $i$, while $\phi_i(t)$ is the
memory field which is updated with the following deterministic rule:
\begin{equation}
\phi_i(t)=\sigma_i(t)+\delta\phi_i(t-1).
\end{equation}
The real parameter $\delta<1$ is related to the length of the memory, in fact we have
\begin{equation}
 \phi_i(t)=\sum_{\tau=0}^t\delta^{\tau}\sigma_i(t-\tau)=\sum_{\tau=0}^te^{-\tau/\tau_c}\sigma_i(t-\tau),
\end{equation}
with a characteristic time defined by
\begin{equation}
\tau_c=-\frac{1}{\log\delta}.
\end{equation}
The factor $(1-\delta)$ in the second term of (\ref{energy}) takes into account the fact that if $\sigma_i=\sigma$ for an infinite time, then $\phi_i\rightarrow \frac{\sigma}{1-\delta}$. The parameter $j_{mem}$ regulates the relative weight of the interaction between spins and memory field with respect to nearest-neighbour interaction. In the following we consider the case $j_{mem}=1$.
 If we define
\begin{equation}
h_i(t)=\sum_j\sigma_j(t)+(1-\delta)\phi_i(t),
\end{equation}
we have that the transition probability $p_i(t)$ for the spin $\sigma_i$ at time $t$ is
\begin{equation}
\label{transition}
    p_i(t)=\frac{e^{\beta h_i(t)}}{e^{\beta h_i(t)}+e^{-\beta h_i(t)}},
\end{equation}
where $\beta$ is the inverse temperature and we set the Boltzmann's constant to one. The update rule is given by
the heat bath dynamics (HBD): the spin at time $t$ is oriented according to the local field by
\begin{equation}
\label{HBD} \sigma_i(t+1)=sgn(p_i(t)-z_i(t)),
\end{equation}
where $z_i(t)$ are independent random numbers uniformly distributed between zero and one.\\
 In a two dimensional square lattice, in presence of memory,
domains stop growing at zero temperature because of the pinning action of corner spins (i.e. spins with two up and two down neighbors). In fact in this situation moves which decrease, leave
unchanged or increase the energy are accepted with probability $1$,$\frac{1}{2}$ or $0$ respectively. Without memory a corner spin  can flip with probability
$\frac{1}{2}$, while in presence of memory they can have an energy smaller than zero and then stop the ordering dynamics. At low but finite temperature, however, we have a
small but finite probability for such a spin to flip even in presence of memory. This is a situation similar to the one of the RFIM, where spins are coupled to a random field and the growth process is slowed down by pinning effects. The memory field is however self generating and correlated in space as the local magnetization, while in the RFIM the field is given at the beginning of the evolution and usually has no spatial correlations.
In order to understand what kind of behavior is induced by the memory,  we numerically studied growth and persistence properties of this system at finite temperature.
\section{Growth and persistence properties}

\begin{center}
\begin{figure}
\begin{center}
\includegraphics[width=14cm]{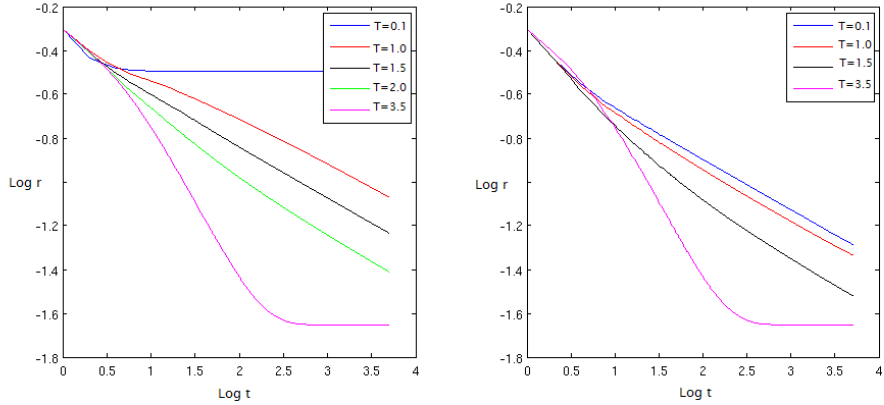}
\caption{\footnotesize{\textit{Fraction of persistent spins as a function of time at various
temperatures.  Left panel: Ising model with memory. Right panel: Ising model. }}}\label{fig:fig1}
\end{center}
\end{figure}
\end{center}

\begin{center}
\begin{figure}
\begin{center}
\includegraphics[width=14cm]{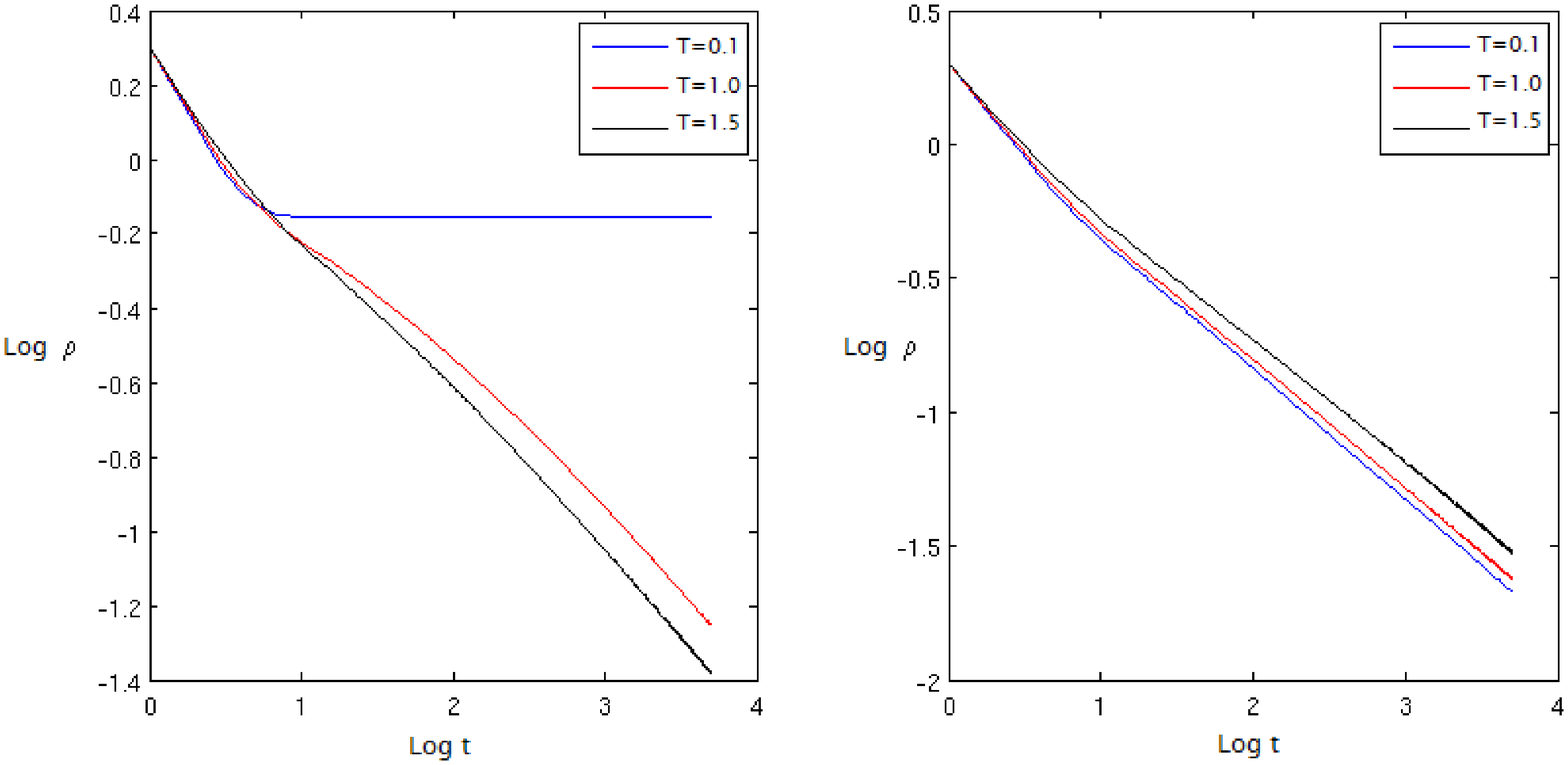}
\caption{\footnotesize{\textit{Density of interfaces as a function of time at various
temperatures. Left panel: Ising model with memory. Right panel: Ising model.}}}\label{fig:fig2}
\end{center}
\end{figure}
\end{center}

\begin{center}
\begin{figure}
\begin{center}
\includegraphics[width=7cm]{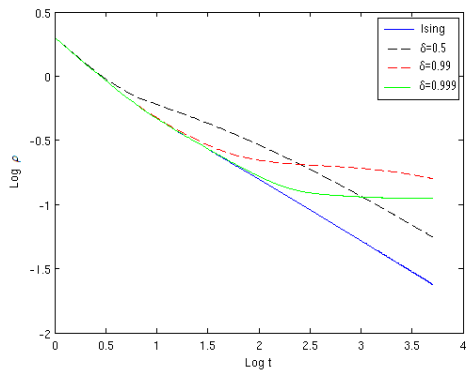}
\caption{\footnotesize {\textit{Ising model with memory: density of interfaces as a function of time at various
memory length.} }}\label{fig:densitydelta}
\end{center}
\end{figure}
\end{center}

\begin{center}
\begin{figure}
\begin{center}
\includegraphics[width=7cm]{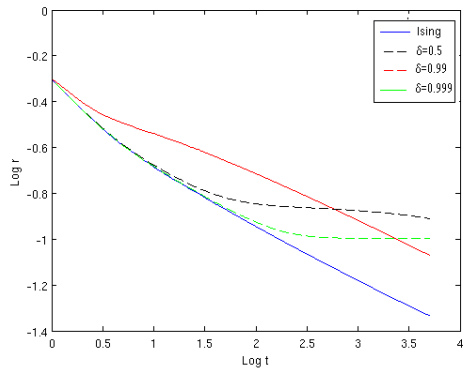}
\caption{\footnotesize {\textit{Ising model with memory: fraction of persistent spins as a function of time at various
memory length.}}}\label{fig:persistentidelta}
\end{center}
\end{figure}
\end{center}

For computing the fraction of persistent spins and the density of interfaces, we used the method introduced by Derrida \cite{Der97}. The idea is to compare the
system (A) where coarsening takes place with a copy of the system (B) submitted to the same noise but with
ordered initial conditions (all spins up). In this way we can distinguish between spin-flips due to the
coarsening process and spin-flips due to the thermal (equilibrium) fluctuations within the bulk of domains. With
this set-up one can measure the density of interfaces as
\begin{equation}
\label {den} \rho=\frac{1}{N}(\sum_{i<j}\sigma_{i}^{A}\sigma_j^{A}-\sum_{i<j}\sigma_{i}^{B}\sigma_j^{B}).
\end{equation}
In the scaling regime this quantity is related to the characteristic length $L(t)$ by
\begin{equation}
\rho\sim L^{-1}(t).
\end{equation}
For the Ising model  this means that
\begin{equation}
\rho\sim t^{-1/2}.
\end{equation}
For the fraction of persistent spins we have
instead
\begin{equation}
\label{persistent}
    r(t)=\frac{1}{N}\sum_{i=1}^N\prod_{0\leq t'\leq t}\left(1+\sigma_i^{(A)}\sigma_i^{(B)}\right).
\end{equation}
Notice that if we prepare the system B initially in the configuration with all the spins up, with the above
definition $r(t)$ is two times the fraction of persistent up spins. One can also improve this calculation
introducing a third copy of the system starting in the all down configuration \cite{Hin97}. For the Ising model at long times and below the critical temperature, the
fraction of persistent spins has the power law decay
\begin{equation}
r(t)\sim t^{-\theta},
\end{equation}
with $\theta\sim 0.22$. This exponent seems to be temperature independent below $T_c$ (\cite{Der97},
\cite{Hin97}), but for $T=T_c$ a faster decay is observed. Above $T_c$ the two systems become quickly identical
and $r(t)$ saturates to a constant value. \\
We repeated the same analysis for the Ising model in presence of memory. We simulated two systems (A and B) of
$1000$ x $1000$ spins with helical boundary conditions. The systems are prepared according to
\begin{equation}
\label{CI} \sigma_i^{A}(0)=sgn\left[\frac{1}{2}-z_i^{(0)}\right],\:\sigma_i^{B}(0)=1,
\end{equation}
where $z_i^{(0)}$ are random numbers between zero and one. The quantities $r(t)$ and $\rho(t)$ are measured up to $5000$ time
steps (in a time step we propose in average a spin-flip for each spin) and averaged over $50$ independent runs.
We made measurements for different values of temperature and memory length (Figures
\ref{fig:fig1},\ref{fig:fig2},\ref{fig:densitydelta},\ref{fig:persistentidelta}). In Figure \ref{fig:fig1} we
can see the fraction of persistent spins as a function of time plotted for different values of temperature with
$\delta=0.5$.

We can identify a regime of temperatures in which the fraction of persistent spins displays, after a transient
time, a power low decay $r(t)\sim t^{-\theta}$ with an exponent which is consistent with the one of the Ising model. Then there is the high temperature regime in which, as for
the standard Ising model, $r(t)$ saturates to a constant value. We observe the power law decay for
the fraction of persistent spins up to temperatures greater than $T_c$, the critical temperature of the Ising
model. We also observe  that $r(t)$ develops a plateau as temperature is lowered
that leads to an (apparent) arrest of coarsening within the simulation time.  \\
For the density of interfaces we have a similar scenario (Figure \ref{fig:fig2}). There is a regime of
temperature in which we observe the scaling behavior of the Ising model, then at lower
temperatures there is an apparent arrest of coarsening.
In Figures  \ref{fig:densitydelta} and  \ref{fig:persistentidelta} we see the behavior of $\rho(t)$ and $r(t)$ for different values of the parameter $\delta$ at fixed temperature.
We observe that at early times, when the memory has still to grow, all the curves start together and the system with memory is not different from the Ising case. Later, when the memory is strong enough, all the curves deviate from the one which represents the Ising model. The time of this deviation depends on the memory length . The longer is the memory, the longer it takes for the system in order to be influenced by it, so the time at which the system deviates from the Ising-like behavior grows with $\delta$.  On the other hand, we see that both the interface density and the fraction of persistent spins develop a plateau as $\delta$ increases, as it happens as temperature is lowered.
The energy barriers due to the presence of the memory
that the system has to overcome in order to evolve become higher the lower the temperature is and the longer the memory is.  Notice that these energy barriers are due to local defects.  A corner spin with saturated memory has in fact a finite probability to flip and to resist in the new state long enough for its memory to change sign. So the dynamics of corner spins is the same as for the Ising model with a different characteristic time. We then expect coarsening to take place for long enough time with the same growth laws of the Ising model \cite{Lai88}.  This is certainly true for the one dimensional case, where domain walls can be considered as random walkers. The memory then introduces a time dependence in the diffusion constant. To characterized the behavior in two dimensions we need to study
the system for very long times.

\section{Urn model}
\begin{center}
\begin{figure}
\begin{center}
\includegraphics[width=7cm]{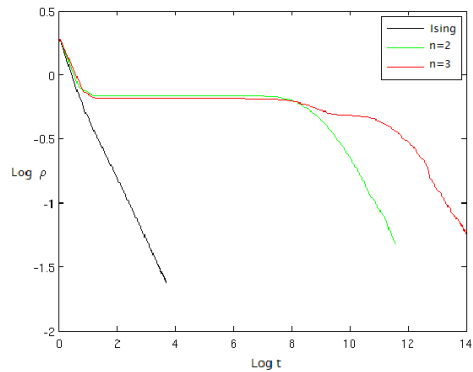}
\caption{\footnotesize{\textit{Urn model, density of interfaces (single runs). }}} \label{fig:urn}
 \end{center}
\end{figure}
\end{center}

At low temperatures
the system does not change state for most of the time, because the acceptance ratio of the Monte Carlo algorithm
is very small. To overcome this problem, a more efficient algorithm is the n-fold
way introduced by Bortz Kalos and Lebowitz (\cite{Bor75}, \cite{Nov01}). This algorithm uses spin classes: every
spin belongs to a class depending on its orientation and on the orientations of its neighbors. The energy of
every spin in a class is the same, so the probability of flipping any spin in a given class is the same. The
idea is to choose a spin, to flip the spin and to update the time according to the probability of having a
spin-flip. In order to apply a similar approach to our problem, it's convenient to consider a memory term which
is slightly different from the one that we have considered up to now. We imagine for each site an urn that
contains $n$ balls, the balls can be of kind $+1$ or $-1$. The memory term that we put in the transition
probability is given by
\begin{equation}
\phi(t)=n_+(t)-n_-(t),
\end{equation}
where $n_+(t)$ ($n_-(t)$) is the number of $+$ ($-$) balls at time $t$.  At each step we choose at random one of
the balls and replace it with a ball of kind $\sigma$. In this way we have for the memory of spin $i$ that

\begin{equation}
\langle\phi_i(t)\rangle=\phi_i(t-1)\left(1-\frac{1}{n}\right)+\sigma_i(t),
\end{equation}
which, on average, is the same behavior of our previous model with the identification
\begin{equation}
\delta\leftrightarrow 1-\frac{1}{n}.
\end{equation}
In Figure \ref{fig:urn} we can see the results for a system with $n=2$ and $n=3$.
At low temperature domains stop growing because of the local barrier introduced by memory. However, there is a characteristic timescale at which they start growing again with the same scaling behavior of the Ising model. This characteristic timescale depends, as usual, on the temperature, since the probability of flipping a corner spin which is parallel to its (saturated) memory is proportional to $e^{-2\beta}$.  We also observe that the characteristic timescale depends on the memory length. We can explain this fact with the following argument. Let us consider a corner spin with saturated memory. We can say that a spinflip is "successful" if the spin flips and the memory changes sign so to support the new state of the spin. The longer is the memory, the longer it takes for a spin in order to change the sign of its memory. The spin has to flip and to persist in its new state long enough for its memory to change sign.

\section{Correlation function}
\begin{center}
\begin{figure}
\begin{center}
\includegraphics[width=14cm]{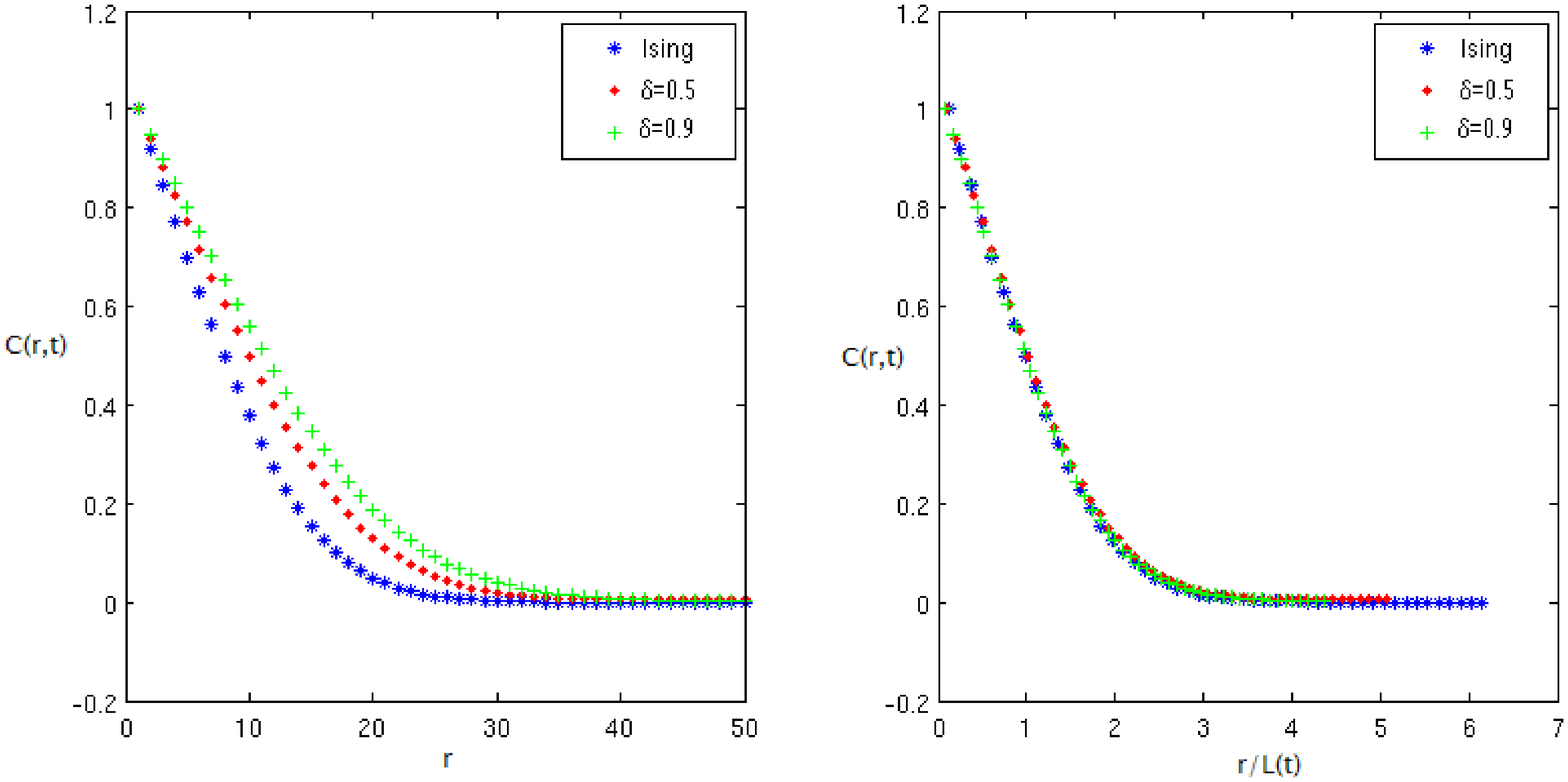}
\caption{\begin{footnotesize} \textit{Collapse of the correlation function: Ising $t=100$, $\delta=0.5$ $t=1000$, $\delta=0.9$ $t=4000$.}\end{footnotesize}} \label{fig:corr}
 \end{center}
\end{figure}
\end{center}

\begin{center}
\begin{figure}
\begin{center}
\includegraphics[width=14cm]{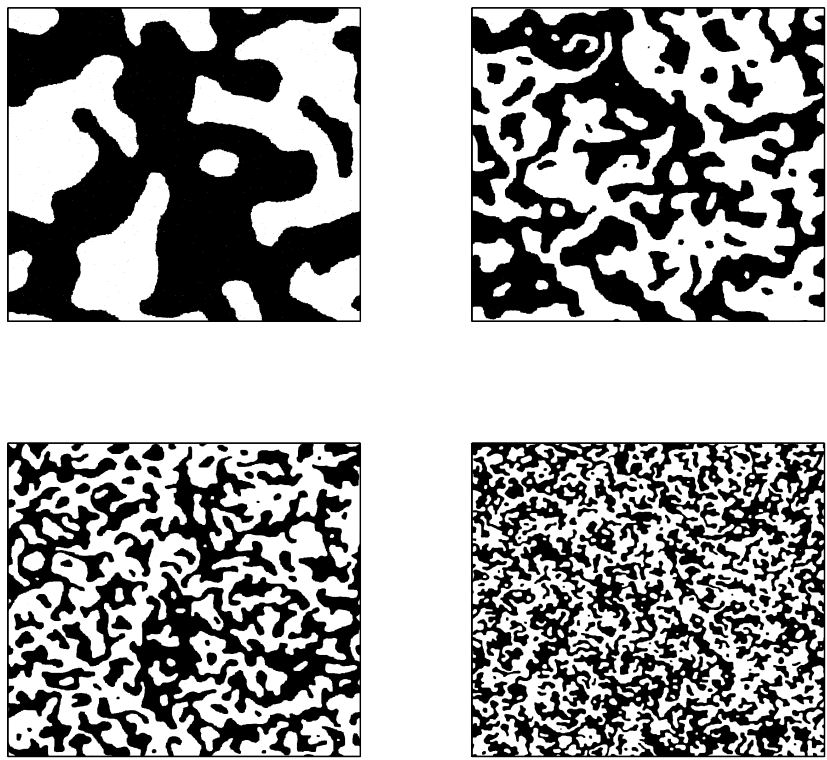}
\caption{\begin{footnotesize} \textit{Snapshot of the system for different values of $\delta$ ($t=5000$): upper left panel Ising
model, upper right panel $\delta=0.5$, lower left panel $\delta=0.9$, lower right panel $\delta=0.99$.}\end{footnotesize}} \label{fig:snap}
 \end{center}
\end{figure}
\end{center}

The existence of a scaling regime characterized by a single length scale $L(t)$ reflects on the structure of the correlation functions. For example the pair correlation function has the form
\begin{equation}
    C(r,t)=f\left( \frac{r}{L(t)}\right) .
\end{equation}
The Ising model and the RFIM have the same scaling functions \cite{Rao93}, which means that the two models
belong to the same universality class. In Figure \ref{fig:corr} we show  the correlation function for different
values of memory as a function of $r/L(t)$.  The characteristic length  $L(t)$ has been measured as the distance
over which the correlation function falls to half its maximum value. We observe a good collapse of the curves,
meaning that the universality class is the same of the Ising model and of the RFIM. From the behavior of the
correlation function we can understand that the energy barriers introduced by the memory can lead to a delay of the domain growth, but that in the scaling regime the morphological structure of domain patterns is the same as that of the
Ising model, as we can see from Figure \ref{fig:snap}.

\section{Conclusions}
We considered a kinetic Ising model in which we introduced a field which breaks detailed balance. This field has a natural interpretation in terms of memory of spins and its introduction can be interesting for possible applications in the context of opinion dynamics. The memory field is self-generating and correlated in space as the local magnetization. We characterized growth and persistence properties for the two dimensional case in a square lattice. At zero temperature the system freezes because of the pinning action induced by memory. At finite temperature domains have to overcome energy barriers in order to grow. This leads to an apparent arrest of growth at low temperature and/or long memories. Since the energy barriers are due to local defects, however,  coarsening takes place at long enough time with the same asymptotic laws of the Ising model. The characteristic timescale depends both on the temperature and on the memory length. The two point rescaled correlation functions  is (for long times) the same of the Ising model and of the RFIM, indicating that the dynamical universality class is the same. 

\end{document}